\documentclass{aa} 

\usepackage{graphicx}
\usepackage{txfonts}
\def\fdeg{\hbox{$^\circ$}}
\def\degr{^\circ}

\newcommand{\BP}{G_{BP}}
\newcommand{\RP}{G_{RP}}
\newcommand{\GKO}{(G-K)_0}

\begin{document}

   \title{Updated {\it Gaia}-2MASS 3D maps of Galactic interstellar dust}


   \author{R. Lallement
          \inst{1}
\and
          J.L. Vergely\inst{2}
          \and
          C. Babusiaux\inst{3}
\and
          N.L.J. Cox\inst{2}
          }
   \institute{GEPI, Observatoire de Paris, PSL University, CNRS,  5 Place Jules Janssen, 92190 Meudon, France\\
              \email{rosine.lallement@obspm.fr}
\and
   ACRI-ST, 260 route du Pin Montard, 06904, Sophia Antipolis, France             \and
             Univ. Grenoble Alpes, CNRS, IPAG, 38000 Grenoble, France
             }

\date{Received ; accepted }

 
\abstract
   {} 
   {Three-dimensional (3D) maps of Galactic interstellar dust are a tool for a wide range of uses. We aim to construct 3D maps of dust extinction in the Local Arm and surrounding regions.}
   {To do this, {\it Gaia} EDR3 photometric data were combined with 2MASS measurements to derive extinction towards stars with accurate photometry and relative uncertainties on EDR3 parallaxes of less than 20\%. We applied our hierarchical inversion algorithm adapted to inhomogeneous spatial distributions of target stars to this catalogue of individual extinctions.} 
{We present the updated 3D dust extinction distribution and provide an estimate of the error on integrated extinctions from the Sun to each area in the 3D map. The full computational area is similar to the one of the previous DR2 map, that is to say with a 6 $x$ 6 $x$ 0.8 kpc$^{3}$ volume around the Sun.  Due to the addition of fainter target stars, the volume in which the clouds can be reconstructed has increased. Due to the improved accuracy of the parallaxes and photometric data in EDR3, extinctions among neighbouring targets are more consistent, allowing one to reach an increased contrast in the dense areas, while cavity contours are more regular. We show several comparisons with recent results on dust and star distributions. The wavy pattern around the Plane of the dust concentrations is better seen and exists over large regions. Its mean vertical peak-to-peak amplitude is of the order of 300 pc; interestingly, it is similar to the vertical period of the spectacular snail-shaped stellar kinematical pattern discovered in {\it Gaia} data.}
{The {\it Gaia} EDR3 catalogue allows for a significant improvement of the extinction maps to be made, both in extent and quality. The hierarchical technique confirms its efficiency in the inversion of massive datasets. Future comparisons between 3D maps of interstellar matter and stellar distributions may help to understand which mergers or internal perturbations have shaped the Galaxy within the first 3 kpc.}


  


   \maketitle
%

\section{Introduction}

Three-dimensional (3D) maps of extinction by the Galactic interstellar (IS) dust are a useful tool in many respects. Among the main applications, they allow de-reddening of spectra to provide intrinsic properties of observed objects, they provide information on the foreground, environment, or background for specific Galactic targets, and they can be used in conjunction with stellar populations and gas distributions to infer the past history of the Milky Way. In the context of interstellar matter studies, they can be used in conjunction with other absorption data and/or multi-wavelength interstellar emission measurements, including polarised emission, and they provide a frame for models of light or particle propagation. Three-dimensional dust maps are constructed through the synthesis of estimated starlight extinctions and they have started to develop rapidly during the last ten years with the advent of massive stellar surveys. Their development is considerably boosted by the {\it Gaia} mission data releases. 

 Three-dimensional dust extinction mapping is based on the tomographic inversion of measured extinction-distance pairs or extinction-distance probability distributions for large numbers of targets distributed in 3D space. What differs among the various 3D reconstructions is the choice of the photometric and(or) spectro-photometric catalogues used to estimate individual extinctions, the source of distances, that is parallax,  photometry, or combined parallax and photometry, and finally the inversion technique. Such choices result in various extents of the mapped volumes. For descriptions of recently constructed maps, readers can refer to \cite{Sale18,Green19,Lallement19,Chen19,Rezaei20,Chen20,Leike20,Hottier20} and \cite{Guo21}. 

One of the main difficulties of 3D IS dust extinction tomography is the multi-scale structure of the interstellar medium, from small dense cores to extended warm clouds. Reconstructing the cores evidently requires a high concentration of targets in their areas. This is also connected to the second difficulty, the lack of homogeneity of the target spatial distribution. In general, the number density of targets decreases with increasing distance due to observational limitations. To adapt to the multi-scale structure, several inversions allowed for a large spectrum of cloud sizes. The corresponding, detailed maps were computed  based on Gaussian random processes or techniques derived from information field theory \citep{Sale18,Rezaei20,Leike20}.  The present work uses a different technique based on iterations of inversions at increasing resolutions. This technique has been previously used to compute a dust map based on {\it Gaia} Data Release (DR2) and 2MASS photometric data combined with {\it Gaia} DR2 parallaxes. The basic principles of the 3D inversion have been detailed in \cite{Vergely01} and the new hierarchical technique has been described in \cite{Lallement19}.  Along this iterative process, the inversion is restricted to regions of 3D space with a target density high enough to achieve the current resolution, that is their total volume is gradually decreasing and there is a simultaneous increase in the volumes where the retained solution is the solution of the previous inversion. This allowed us to adapt the spatial resolution to the concentration of the targets at each location. The advantage of this hierarchical technique is the limited cost in computational time and the possibility to increase the physical dimensions of the maps while avoiding creating unreal structures due to the scarcity of targets. 

The {\it Gaia} Early Data Release 3 \citep[EDR3,][]{GAIACOLL2021} contains a new catalogue of parallaxes and luminosity measurements, with significant improvements with respect to {\it Gaia} DR2, in particular the accuracy of parallaxes has significantly increased and the homogeneity of photometric measurements has thoroughly improved. We made use of the new measurements to produce a larger and higher quality catalogue of individual extinctions to enter the inversion process.

Section 2 describes the selection of EDR3 data and the extinction estimate method. Section 3 recalls the main characteristics of the hierarchical tomographic inversion and how it will be adapted to the new input data. It contains the description of error estimates and potential biases. Section 4 contains comparisons between the map and locations of molecular clouds based on different data, as well as a comparison between the dust distribution and young stars close to the mid-plane. It describes several features of the new 3D map. Section 5 contains a summary and discusses mapping perspectives.

\begin{figure}[t]
\centering
\includegraphics[width=7cm]{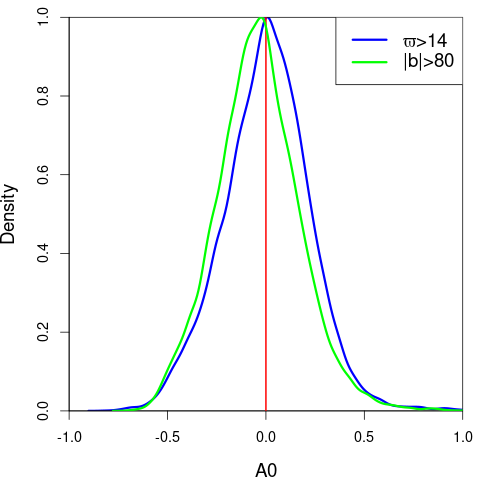}
\caption{A$_{0}$ extinction distribution (mag) in low extinction selections. In blue, the local bubble selection is provided ($\varpi>14$ mas, corresponding to the first 70~pc); and in green, a high latitude selection is shown ($|b|>80\degr$, $|Z|<400$pc).}
\label{fig:lowexttest}
\end{figure}

\begin{figure}[t]
\centering
\includegraphics[width=7cm]{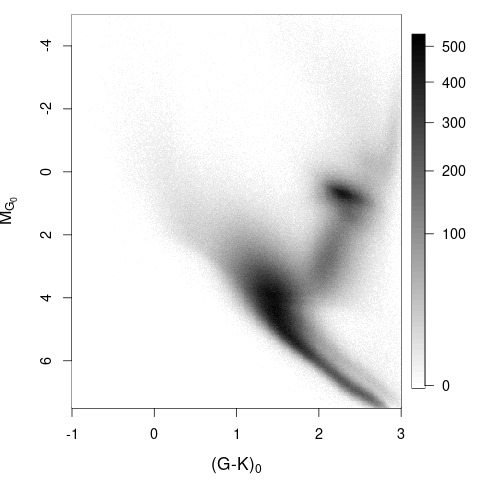}
\caption{De-reddened Hess diagram. The greyscale corresponds to the square root of the stellar density. The red clump is the most prominent feature whose shape can be compared to Fig. 2 of \cite{RuizDern17}.}
\label{fig:hrdtest}
\end{figure}

\begin{figure*}
 \centering
  \includegraphics[width=0.46\hsize]{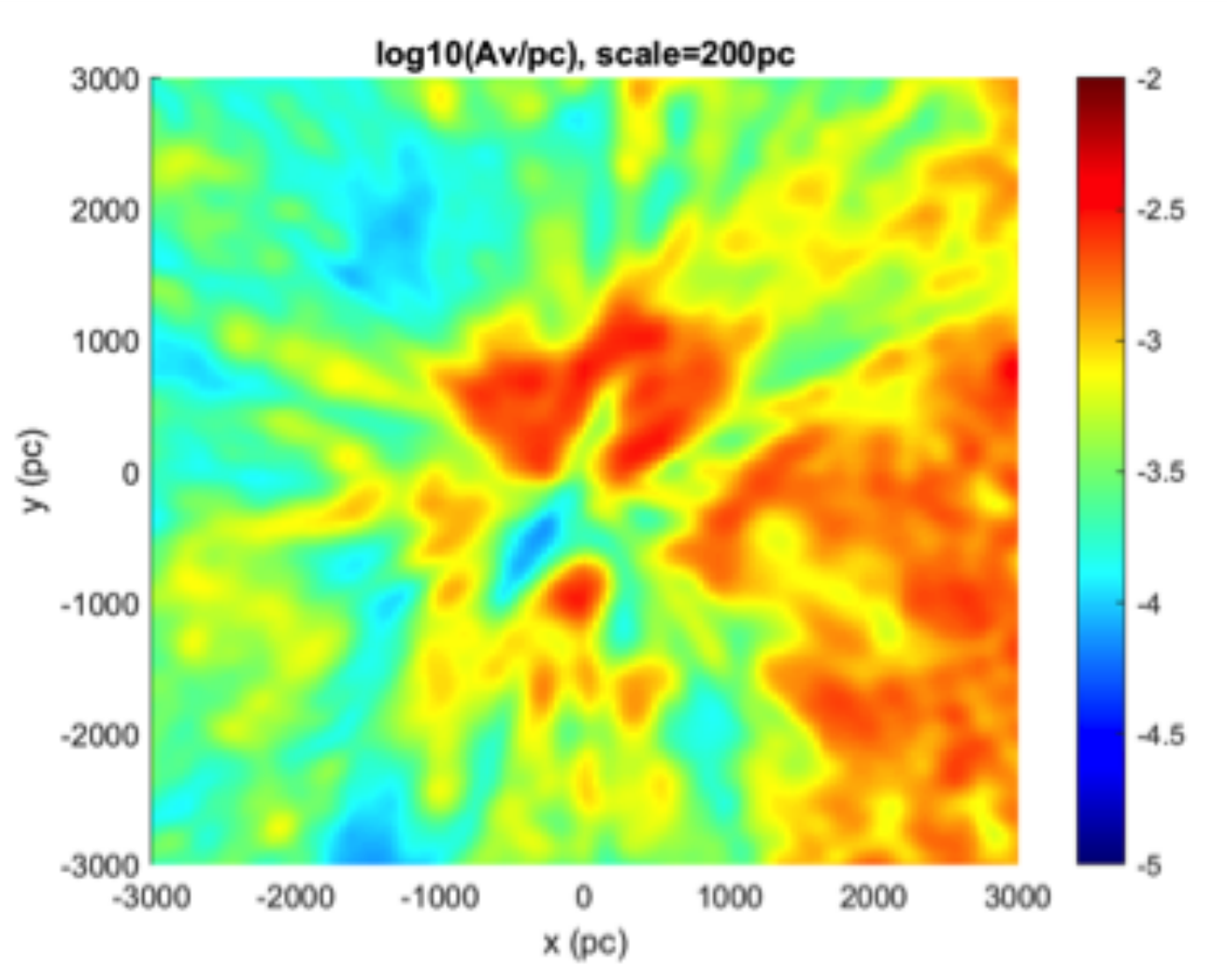}
  \includegraphics[width=0.46\hsize]{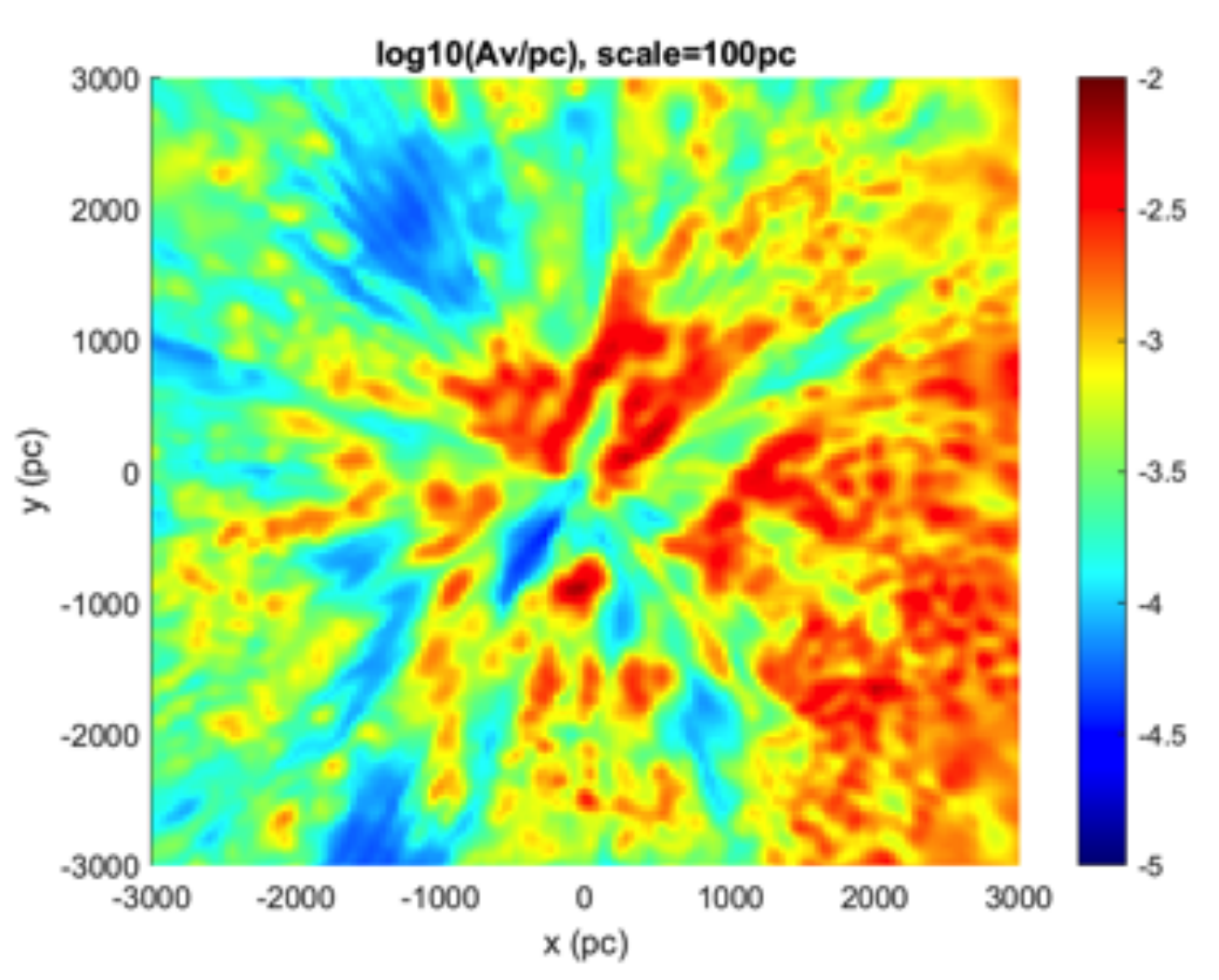}
   \includegraphics[width=0.46\hsize]{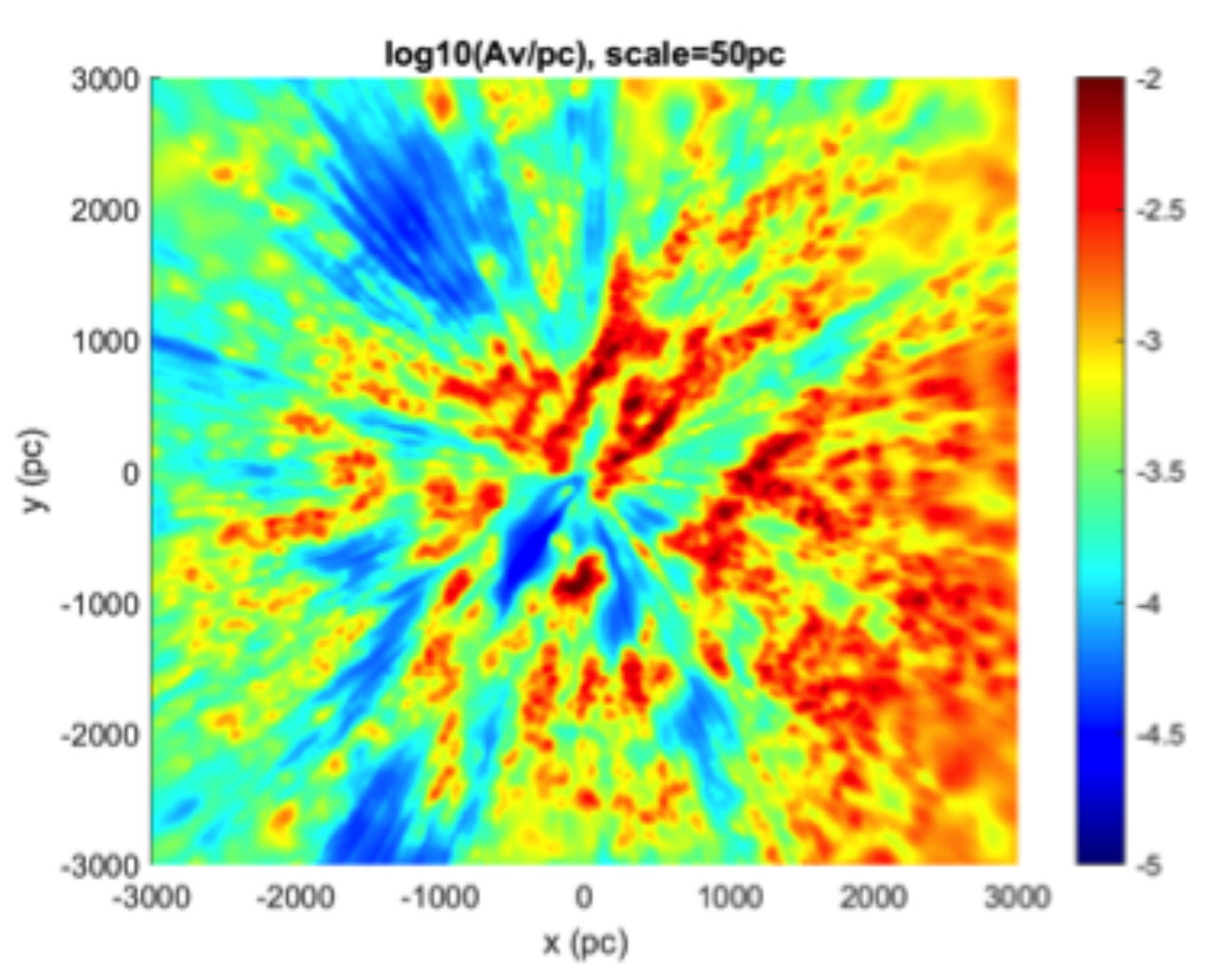}
    \includegraphics[width=0.46\hsize]{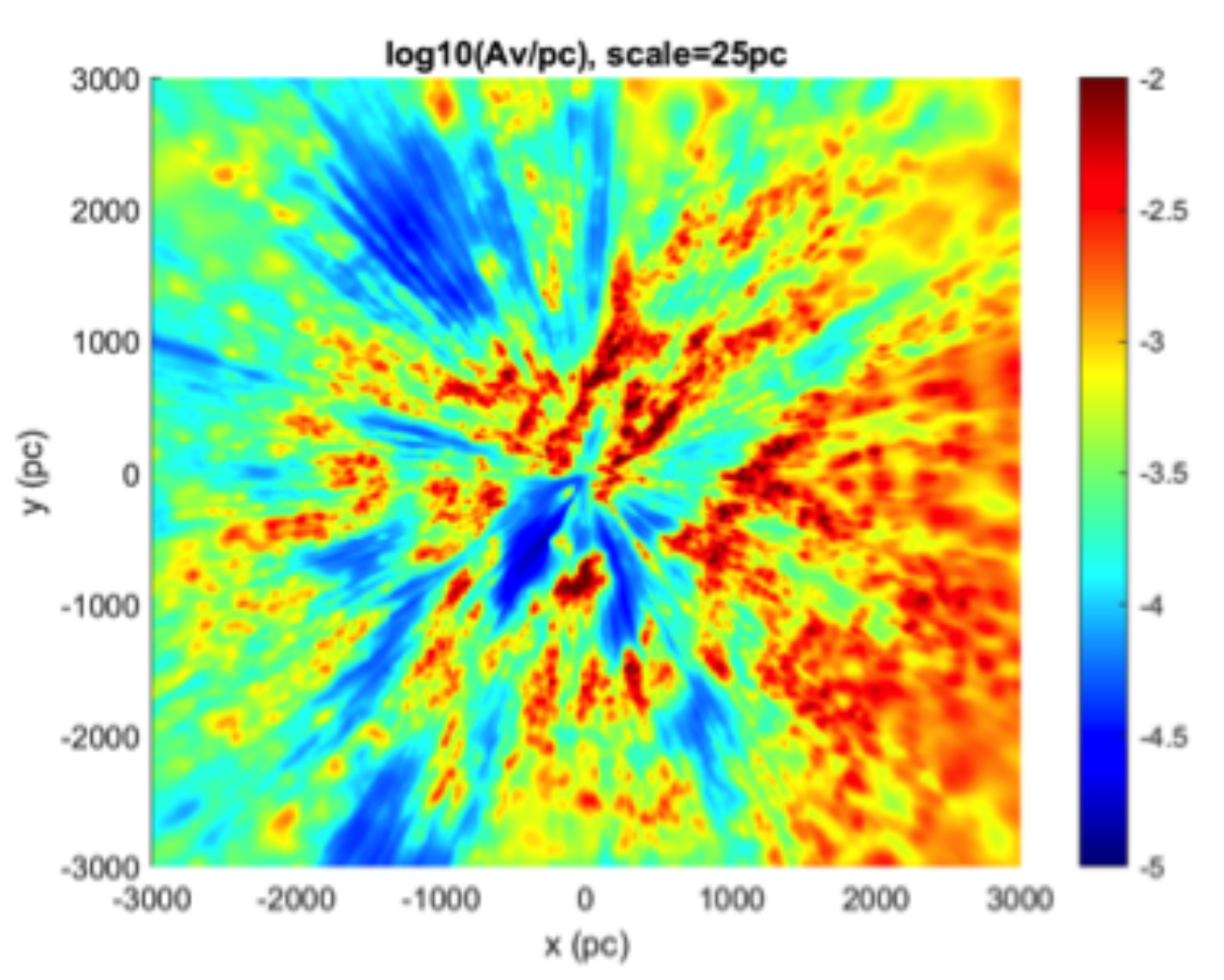}
  \caption{Extinction densities in the mid-plane for four scales: 200pc, 100pc, 50pc, and 25pc}
  \label{fig:hierar}%
    \end{figure*}

\begin{figure*}
 \centering
  \includegraphics[width=0.98\hsize]{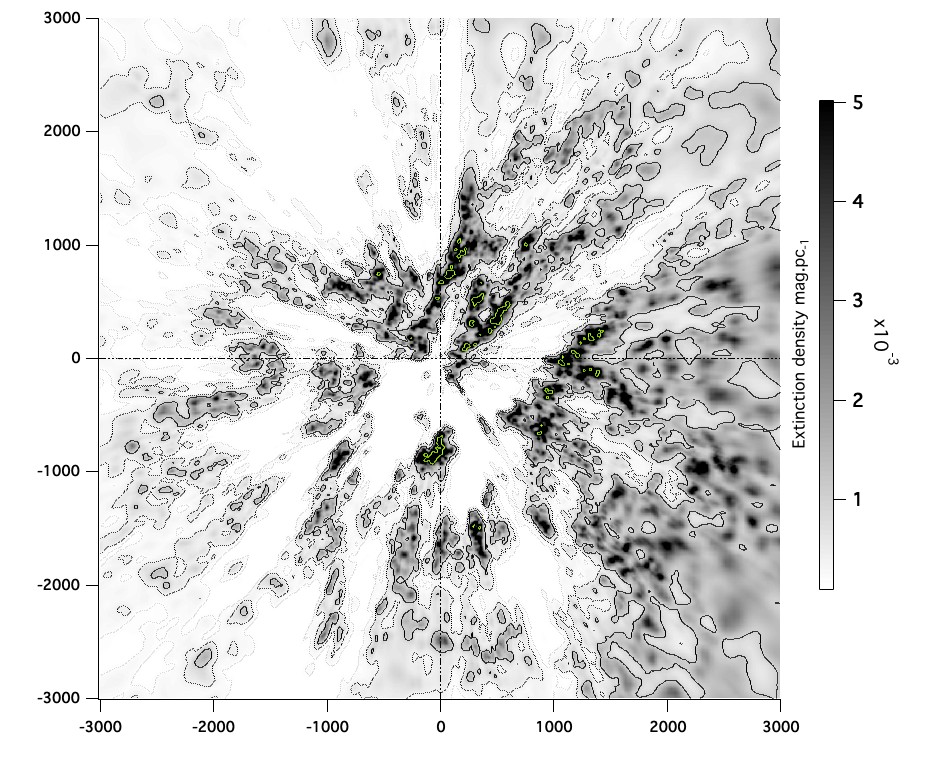}
 \caption{Differential extinction (or extinction density) distribution, here in the plane parallel to the Galactic Plane and containing the Sun. Units are in parsecs. The Sun is at (0,0) and the Galactic Centre direction is to the right. The green isocontours surround the densest areas at the level of 0.01 $mag\cdot pc^{-1}$. Other contours correspond to 0.00018, 0.0005, and 0.001 $mag\cdot pc^{-1}$.}
  \label{galplane}%
    \end{figure*}

\begin{figure*}
 \centering
  \includegraphics[width=0.88\hsize]{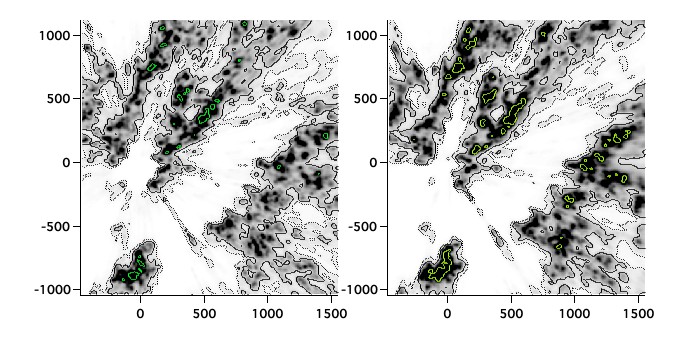}
 \caption{Comparison between the new eDR3 inverted extinction density and the previous result using {\it Gaia} DR2. Here we focus on the fraction of the 3D map containing the densest regions of the Local Arm, the so-called 'split', and Sagittarius. {\bf Left}: {\it Gaia} DR2 map (\cite{Lallement19}. {\bf Right}: {\it Gaia} EDR3 map. The green contours mark the densest areas, with the same level of 0.01 $mag\cdot pc^{-1}$ in the two maps. We note the increased fraction of areas above this value, and, on the contrary, the disappearance of faint artefacts within the cavities, which are both due to reduced uncertainties on distances and estimated extinctions. The local cavity around the Sun is better defined thanks to the addition of nearby, bright targets, and it appears narrower.}
  \label{Fig_compar_planes}%
    \end{figure*}
    
    \begin{figure*}
 \centering
  \includegraphics[width=0.86\hsize]{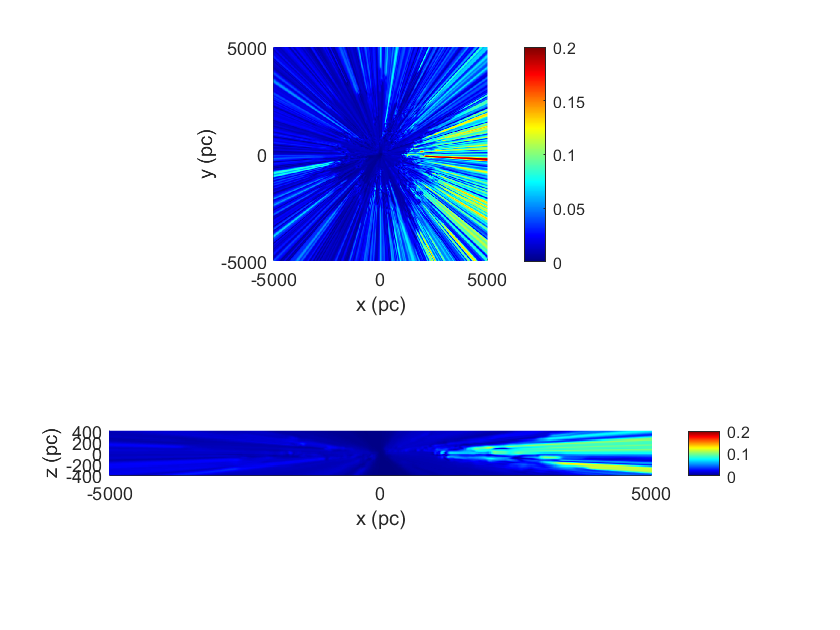}
 \caption{Relative error of the extinction density in the Galactic Plane (top) and in a perpendicular cut (bottom).}
  \label{fig:error2}%
    \end{figure*}

\begin{figure*}
 \centering
  \includegraphics[width=0.86\hsize]{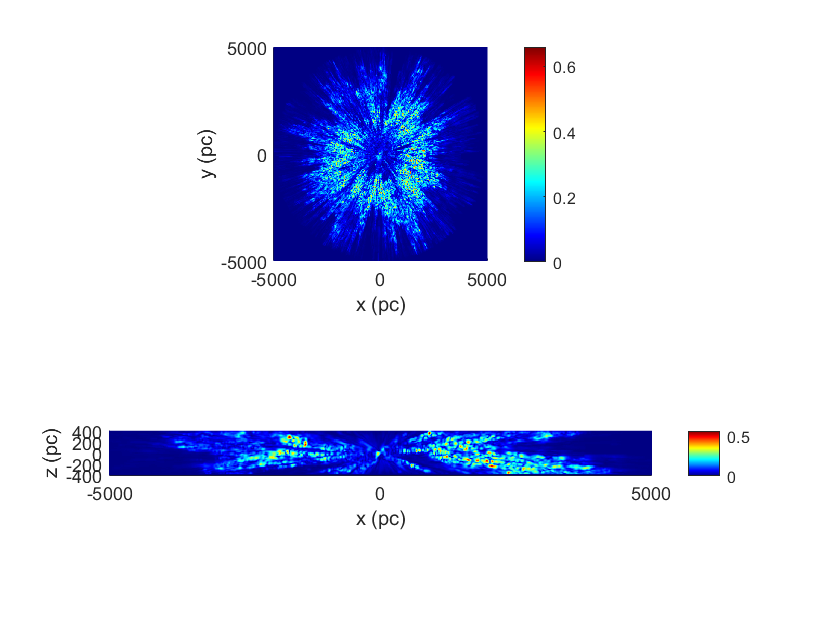}
  \caption{Relative error on the integrated extinction in the Galactic Plane (top) and in a perpendicular cut (bottom). }
  \label{fig:error1}%
    \end{figure*}

\section{Extinction estimates}\label{extinctions}

To derive the extinction of individual stars, we combined the {\it Gaia} EDR3 \citep{GAIACOLL2021} and the 2MASS \citep{2006AJ....131.1163S} photometry. We used the catalogue's cross-match provided within the {\it Gaia} archive. We selected sources with a parallax uncertainty lower than 20\%, astrometric renormalised weight error $\texttt{\small{ruwe}}<1.4$, photometric uncertainties lower than 2\% in $G$, 5\% in $\BP$ and $\RP$, $\BP<20.3$ to avoid biases at low $\BP$ fluxes \citep{Riello21}, and $G>5$ to avoid saturation. The corrected flux excess was used to remove sources affected by crowding following the recipe of \cite{Riello21} at 10$\sigma$. The $G$ photometry for the six-parameter solutions was corrected following \cite{Riello21}. The parallax was corrected from the zero point variation with magnitude, colour, and ecliptic latitude following \cite{Lindegren21}. For 2MASS, we selected stars with 2MASS photometric quality flag \texttt{\small{AAA}} and photometric uncertainties lower than 0.05 mag. Only stars with an absolute magnitude $M_G<7.5$ were selected to avoid the near-infrared discontinuity of the colour-colour relation for dwarfs redder than $G-K>3$. A first selection on this criteria was done at the query level assuming no extinction.

Similarly to  \cite{Lallement19}, we derived intrinsic colour-colour relations using low extinction stars. Here, we selected stars having $A_0<0.05$~mag using the 3D extinction map of \cite{Lallement19} (here and throughout the paper, the subscript 0 means the monochromatic value at 5500\AA). The main difference between our study and that of \cite{Lallement19} is that we went to smaller absolute magnitudes so that we had to derive to separate colour-colour relations: one for the top of the HR diagram ($M_G<4$) and one for the dwarf sequence, with the hot dwarfs being part of both calibrations, ensuring the continuity. For the top of the HR diagram, we selected only stars within 500~pc of the Galactic Plane to avoid being dominated by metal-poor giants.
A 7 degree monotonic polynomial\footnote{R package MonoPoly} was adjusted to each colour-colour relation $G-X$ as a function of $G-K$, removing strong outliers (ten times the median absolute deviation (MAD) of the residuals) one by one. The colour range of our calibration is $-1.0<\GKO<3.7$ for the top of the HR diagram and  $-1.0<\GKO<3.0$ for the dwarfs. It is important to note that the colour range is smaller than in \cite{Lallement19} due to the selection of stars near the Galactic Plane. 

The extinction coefficients used are provided in the {\it Gaia} auxiliary data pages\footnote{https://www.cosmos.esa.int/web/gaia/edr3-extinction-law}. They have been derived similarly to those used in \cite{Lallement19} with the update of the {\it Gaia} passbands from \cite{Riello21} and the extinction law of \cite{Fitzpatrick19}. 

Similarly to  \cite{Lallement19}, for each star the extinction $A_0$ and the intrinsic colour $(G-K_s)_0$ and their associated uncertainties were determined through a maximum likelihood estimation\footnote{R package bbmle}, 

\begin{equation}
\mathcal{L} = \prod_X P(G-X | A_0, (G-K)_0)
\end{equation}
for $X=G_{BP}, G_{RP}, J, H$. To avoid local minima, the following three different initial values were tested: $(G-K)_0=G-K$ (no extinction), as well as $(G-K)_0=1.5$ and $(G-K)_0= (G-K)_\mathrm{max}$ (the maximum colour of the intrinsic colour relation). Both the top of the HR diagram and the dwarf sequence calibrations were tested. We used the intrinsic colour-colour relations to derive $(G-X)_0$ from $\GKO$. Then $P(G-X | A_0, (G-K)_0) = P(G-X | (G-X)_0 + k[A_0,(G-K)_0] A_0)$. We modelled this probability by a Gaussian, quadratically adding the photometric error in the $X$ band and the intrinsic scatter of the colour-colour relation. 
Negative values of $A_0$ were allowed to ensure a Gaussian uncertainty model, which is needed for the inversion method. However no extrapolation of the extinction coefficients was carried out, the negative values of $A_0$ being replaced by 0 to derive the extinction coefficients. 
Typical resulting uncertainties are similar to \cite{Lallement19}, that is 0.3~mag in $A_0$ and 0.2~mag in $\GKO$.
A chi-square test was performed to check the validity of the resulting parameters, removing stars with a p-value limit smaller than 0.05. 
We removed stars with uncertainties on the derived $A_0$ and $\GKO$ higher than 0.5 and 0.4~mag, respectively, as well as outliers for which $A_0$ is not compatible with being positive at 3 sigma. We also removed stars for which the absolute magnitude did not correspond to the colour-colour calibration data used at 3$\sigma$ and outside the calibration range, that is $\GKO>3$.

In contrast to \cite{Lallement19}, the high latitude stars present a small bias towards negative values (Fig.~\ref{fig:lowexttest}). This is due to the fact that we restricted our calibration sample to the rather metal-rich population of the Galactic Plane, leading to a  bias in the metal-poor-dominated population of a high latitude sample. In the de-reddened HRD (Fig.~\ref{fig:hrdtest}), we see that the issue affecting the blue stars has been resolved with EDR3. However a population of stars on the left of the red giant branch is present, which should not be there. This is due to a degeneracy between two different solutions for the maximum likelihood. We suspect that the extinction coefficients we used do not accurately reproduce the behaviour for the red giants in the near-infrared bands. In future work, other spectral energy distributions will be tested to derive the extinction coefficients. We note that those stars were too faint in the $\BP$ band to be in the \cite{Lallement19} sample. Also, it is important to note that those potential biases introduced by these objects are taken into account in the errors (see next section).

More than 35 million location-extinction pairs were obtained from this procedure and may enter the inversion. This is $\simeq$30\% more than for the catalogue based on DR2, and it allowed us  to obtain better constraints on the sources of extinction. 

\begin{figure*}
 \centering
  \includegraphics[width=0.86\hsize]{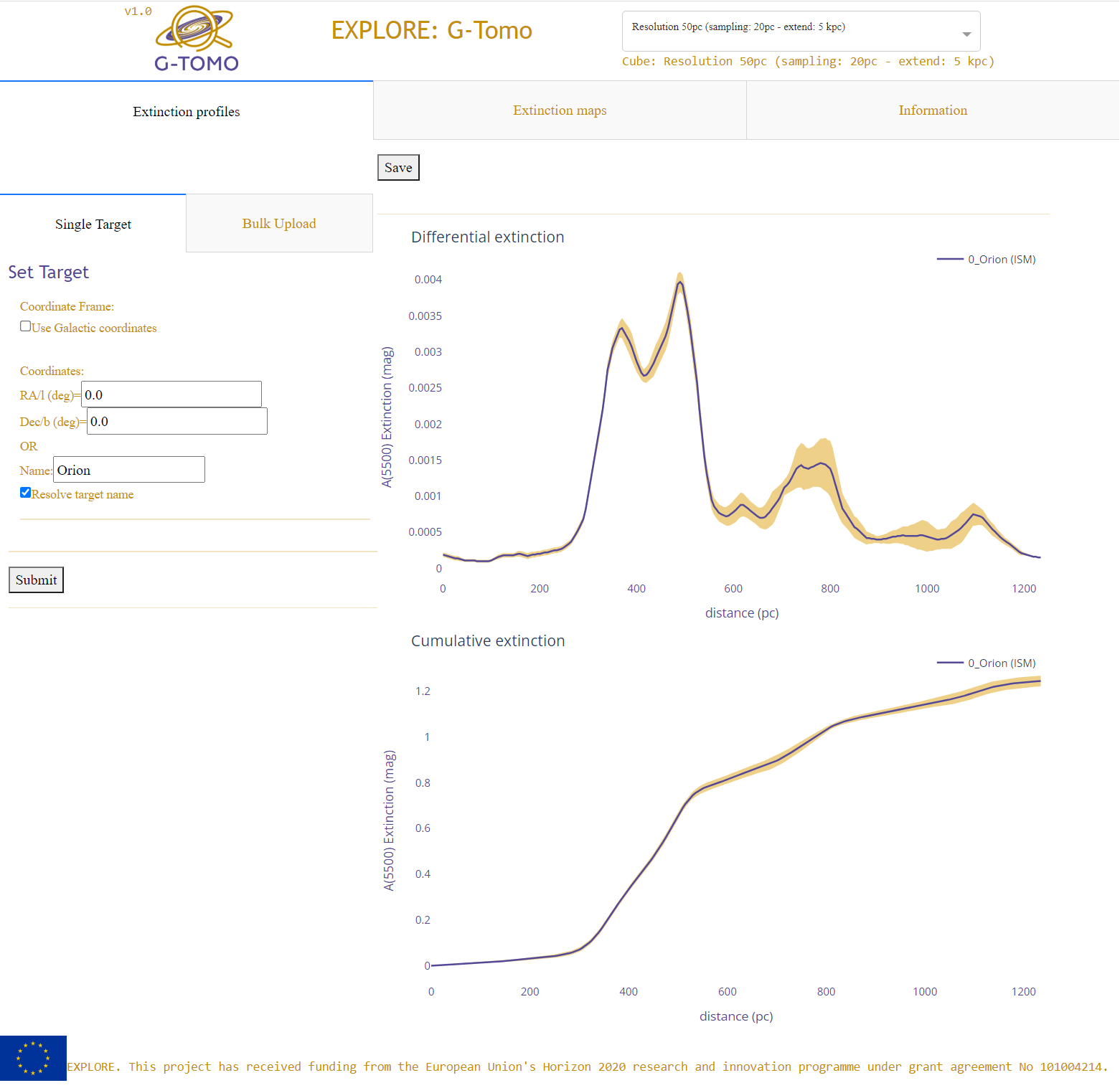}
  \caption{Online tool for extinction and extinction density radial profile: an example.}
  \label{gtomo_prof}%
    \end{figure*}

\begin{figure*}
 \centering
  \includegraphics[width=0.86\hsize]{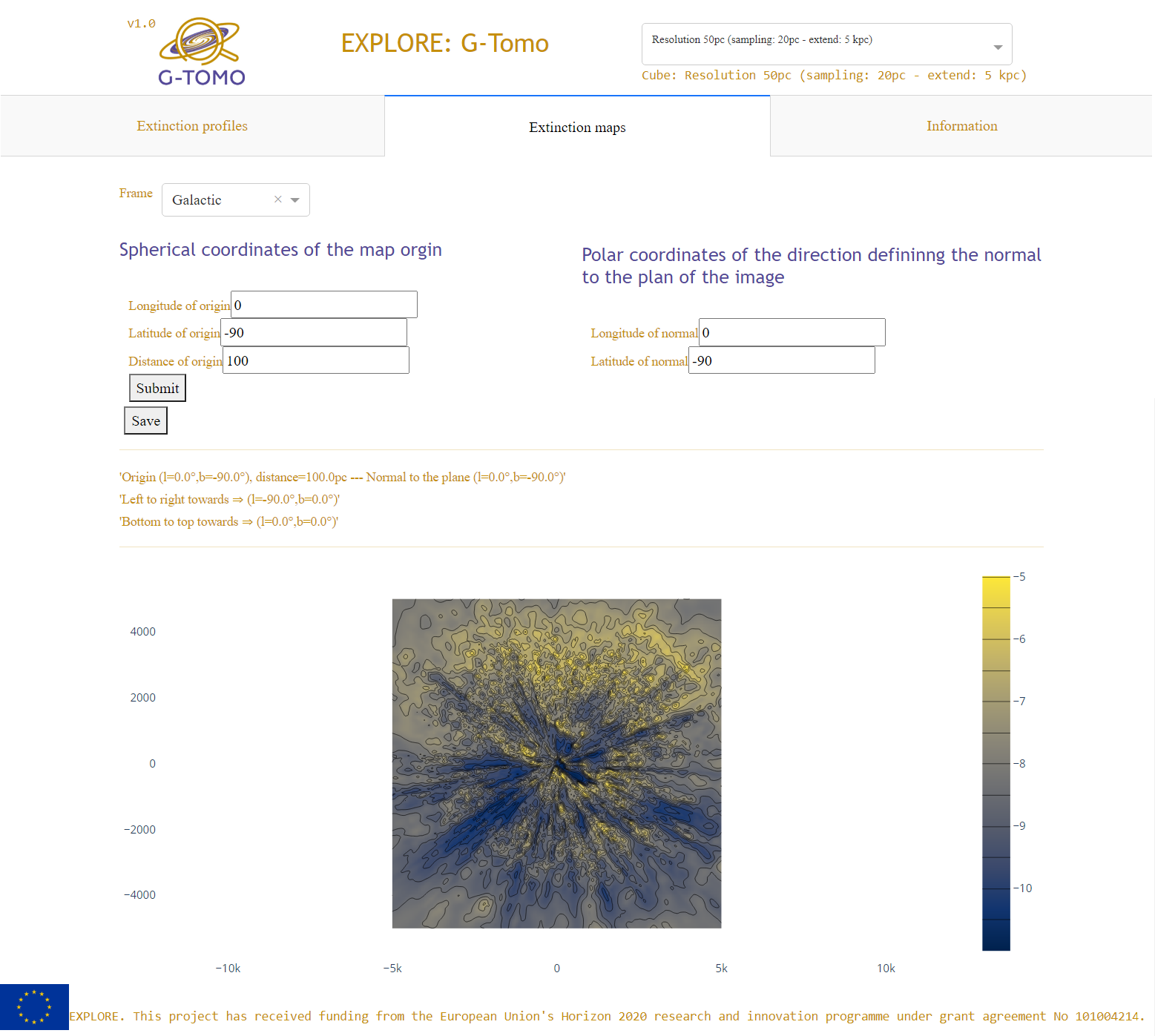}
  \caption{Online tool: example of extinction-density image. Here the normal to the plane of the image is defined by its  Galactic coordinates (0,-90) and  the origin of the image is defined by its distance  from the Sun r=100pc and its Galactic coordinates (0,90). As a result, we show the image of the dust in a plane parallel to the mid-plane and it is displaced by 100 pc to the south. We note that in this example we attempted an inversion up to five kpc.}
  \label{gtomo_plane}%
    \end{figure*}

\section{Hierarchical inversion of individual extinctions and error estimates}\label{inversions}

The 3D tomography algorithm used for inverting the new {\it Gaia} EDR3-2MASS extinction catalogue is similar to the one used for inverting {\it Gaia} DR2-2MASS data \citep{Vergely01,Lallement19}. We estimated the extinction per unit distance $\rho$ = A$_{0}$ / pc (or equivalently the extinction density, a quantity directly related to the dust volume density) in all points in 3D space based on the series of distance-extinction pairs for stars spatially distributed  around the Sun. The multi-scale approach we used consists in estimating the extinction density in an iterative way, starting with large spatial scales and moving towards decreasing scales (or increasing spatial resolution). This method makes it possible to take account of the heterogeneity of the distribution of the lines of sight. Indeed, some areas of the sky are very well covered by data, while others are not; the second situation happens beyond a certain distance or behind dense obscuring clouds.  For each spatial scale, two consecutive types of computations were performed. First, the space was divided into quasi-cubic volumes of a linear size equal to the considered scale divided by two and an estimate of the mean extinction was computed in each of those volumes. Inversions were performed in radial directions and radial extinction density profiles were then discretised and the resulting data points were accumulated into an extinction density cube which was updated direction by direction. Second, the resulting set of average extinctions, associated with the corresponding average distances entered a 3D inversion, using a correlation length equal to the current spatial scale. If, in some of the individual volumes entering the 3D inversion, the number of contributing stars was below a scale-dependent threshold, then this volume was ignored, that is, we did not estimate the extinction density in the vicinity of the barycentre of this volume and kept the value obtained at this location during the previous iteration.  If we had not proceeded in this way, some spatially spread structures
of  extinction which have no physical reality could have appeared in the vicinity of strong extinction areas (see for example the results obtained without the hierarchical process and shown in \cite{Capitanio17}). Moreover, in such a  way, the estimated density structures are spatially more isotropic and the 'fingers' of God' effects are reduced.

Similarly to the algorithm used in Lallement et al. (2019), outliers were detected by means of comparisons with neighbouring stars. Here a 3$\sigma$ selection was applied by comparing individual extinctions with the median of the extinction at each given scale. This means that the outlier detection depends on the scale investigated. At large spatial scales, some high extinction stars may be removed if the majority of the stars are affected by low extinction. However, the same high extinction stars may not be considered as outliers at smaller spatial scales. This strategy avoids the extinction density at large scales to be overestimated as was the case in the past in the context of the first 3D map computations (Vergely et al. 2001). Each 3D inversion is a Bayesian inversion which, for a given scale, uses the 3D solution estimated at the previous scale as a prior. The extinction density model was thus updated at each scale by considering data averaged over smaller and smaller volumes, until a resolution of 25pc was reached (i.e. a sampling of 10pc). Due to the target density threshold mentioned above, the achieved resolution is space-dependent and is better (resp. poorer) in regions of high (resp. low) target density.  In order to illustrate the process, we show the results of iterative inversions in Fig. \ref{fig:hierar} for the 200pc, 100pc, 50pc, and 25pc scales.  What is represented is the computed extinction density distribution in the mid-plane. We note that by mid-plane, we mean the plane containing the Sun which is parallel to the true Galactic mid-plane, that is, we neglected the small distance between the Sun and the true Galactic Plane.  In all figures we use Cartesian coordinates with the Sun at centre X,Y,Z=0,0,0. The X axis is directed to the Galactic Centre, the Y axis is along the direction of rotation, and the Z axis points to the Northern Galactic Pole.  Since the data coverage decreases with distance, for the consecutive scales of 200pc, 100pc, 50pc, and 25pc, the computations were done in cubes of  $20\,{\rm kpc}\,\times\,20\,{\rm kpc}\,\times\,0.8\,{\rm kpc}$, $20\,{\rm kpc}\,\times\,20\,{\rm kpc}\,\times\,0.8\,{\rm kpc}$,  $10\,{\rm kpc}\,\times\,10\,{\rm kpc}\,\times\,0.8\,{\rm kpc}$, and $6\,{\rm kpc}\,\times\,6\,{\rm kpc}\,\times\,0.8\,{\rm kpc}$, respectively.\ However, we show the results in the smallest, final mid-plane, that is a {$6\,{\rm kpc}\,\times\,6\,{\rm kpc}$} plane. The computation time to achieve this 25pc resolution map was about 5 days using a four cores processor.

A frequent request from users of 3D dust maps is the availability of accompanying distributions of errors, most of the time on integrated extinctions or more rarely on extinction densities. As we have warned in previous presentations of maps, our use of a correlation kernel prevents us from reaching precision on the latter quantity because the extinction is forced to be distributed in volumes that can be wider than the actual clouds. On the other hand, it is important to note that errors on integrated extinctions, the measured quantities, do not suffer from the imposed finite resolution as much and they are expected to be much smaller.  Here the error budget could not be computed by means of a classical approach, that is Gaussian error propagation. The main contribution to the error comes from uncertainties on both extinction and distance; however, the target distribution is an important factor too, and the correlation length has an impact on extinction density errors. We have preferred a statistical approach and used errors on target extinctions and distances to build randomly distributed datasets of extinctions and distances around their observed values. Such {noisy} data then entered the hierarchical approach separately which we previously described as providing extinction densities. In such a way, the actual spatial distribution of the targets was taken into account in addition to distance and extinction uncertainties. Here we proceeded in this way five times and we therefore built five cubes of extinction densities. The standard deviation for these five realisations gives an idea of the errors on the extinction density. In parallel, errors on the integrated extinction can also be extracted by calculating, for the five cubes, the extinction integrals from the Sun to any point and by calculating the standard deviation of these integrated extinctions. We note that this error estimate may be affected by the presence of remaining outliers or by the absence of data in certain areas. Moreover, beyond a few kiloparsecs, the computed error is underestimated because the model is drawn towards the prior which is constant in the absence of data. Nevertheless, the order of magnitude of the error may be inferred from the results at a shorter distance. In order to be more conservative, we computed the errors in the case of the 50 pc kernel.

\section{Results}

\subsection{Mapping improvements}

 Fig. \ref{galplane} displays the new EDR3 extinction density distributions in the X,Y plane. The Galactic Centre direction (or X axis) is to the right and the direction of rotation (Y axis) is towards the top of the figure. The size of each pixel is 5pc x 5pc. Units are $mag\cdot pc^{-1}$. Several iso-contours have been drawn to help visualise both dense and faint structures. Compared with the {\it Gaia} DR2-2MASS inverted distribution, there are additional dust structures at large distances that were not recovered previously. This extension is more visible in the first and fourth quadrants due to the existence of numerous clouds from the Sagittarius and Scutum arms, but it is also effective in the two other quadrants, for example we can now detect the prolongation in the second quadrant of the Perseus clouds from the third quadrant. Although distant structures are mapped at lower resolution, as illustrated in Fig. \ref{fig:hierar}, the increased number of targets allows for the 3 kpc x 3 kpc x 0.8 kpc volume for the large structures to be covered entirely. 

 The second main evolution is the increased contrast, that is the larger number of peak densities and, conversely, the increased {vacuity} of the large cavities. To illustrate this, isocontours for the same relatively high value of differential opacity, namely 10$^{-2}$ $mag\cdot pc^{-1}$,  are drawn on both the previous and new map in Fig. \ref{Fig_compar_planes} in a smaller region around the Sun, and they help to visualise the increased number of opaque areas in the new map. Because our minimum kernel is similar in the two maps (25pc), this trend is a sign of better consistency within the dataset, and it reflects the improved accuracy of parallax distances and extinction estimates. Conversely, a careful look at the differences between the two images shows that most of the cavities appear to be emptier, and elongated structures within them that were due to a lack of consistency of the data or distances that were  too uncertain have disappeared in the new map. This is also consistent with the increase in contrast and a sign of a more reliable inversion.  In particular, the 'avenue' devoid of dust between the 'split' and the inner border of Sagittarius looks even more conspicuous and extends up to 2.5 kpc from the Sun in the second quadrant. 
 Finally, we must draw attention to the local cavity (or the 100-200 pc low density, wide area around the Sun). Due to the addition of very nearby target stars, the cavity is more accurately mapped and appears narrower, as can be seen in Fig. \ref{Fig_compar_planes}. 

\subsection{Error estimates}

Orders of magnitudes of errors on the extinction densities (or dust volume densities) are shown in Fig.  \ref{fig:error2}. The performed computations show that the average uncertainties on their amplitudes are around 30 or 40\%  in the main clouds for a 50 pc kernel (Fig. \ref{fig:error2}). In the case of the 25 pc kernel, errors should in principle be smaller; however, we have kept these 50 pc estimates to be more conservative. Such large values are not unexpected, as explained in section \ref{inversions}. They confirm that the maps give an order of magnitude of the dust concentration, but accurate values within the dense clouds are out of reach. We note that, because part of the uncertainties on the extinction density arise from the fact that there are uncertainties on the positions of the clouds (due to parallax errors), the standard deviation of different error realisations increases in regions with a high density, characterised by small clouds. 

Errors on the integrated extinctions are shown in Fig. \ref{fig:error1}. They are of the order of a few percent in most regions, that is to say they are much smaller than the errors on the extinction densities. It may appear surprising since an integrated extinction is a sum of extinctions' densities, but the classical scheme of the square root of the quadratic sum of errors weighted by distances does not apply here since the integrated extinctions are the constrained quantities.

\subsection{Online tools}

The new EDR3 extinction density distribution and corresponding error can be downloaded under the form of a matrix in Cartesian Galactic coordinates centred on the Sun at the EXPLORE website\footnote{ https://explore-platform.eu}. Several online tools are additionally available. The first tool is the computation of cumulative extinction radial profiles for any direction entered in Galactic or equatorial coordinate (ICRS) systems. Both figures and corresponding numerical files can be downloaded. A series of coordinates can also be entered in a single file. For users interested in the distribution of clouds as a function of distance along a line of sight, the computation of the extinction density can be computed and downloaded too. For both tools, the step is 5pc.  Fig. \ref{gtomo_prof} shows an example of both outputs. 

A second tool is the computation of the image of the extinction density distribution in any plane that does or does not contain the Sun for any orientation. The user is asked for the Galactic (or ICRS)  coordinates and distance of the image centre, and the Galactic (or ICRS) coordinates of the normal to the desired plane. To avoid any ambiguity in the orientation of the image, the coordinates of the X and Y axes are indicated. Fig. \ref{gtomo_plane} shows an example of a plane parallel to the mid-plane and displaced by 100 pc to the south (i.e. crossing the Southern Galactic Pole axis at Z=-100pc.).

\section{Comparisons with recent results on dust and stars}

\begin{figure*}
 \centering
 \includegraphics[width=0.88\hsize]{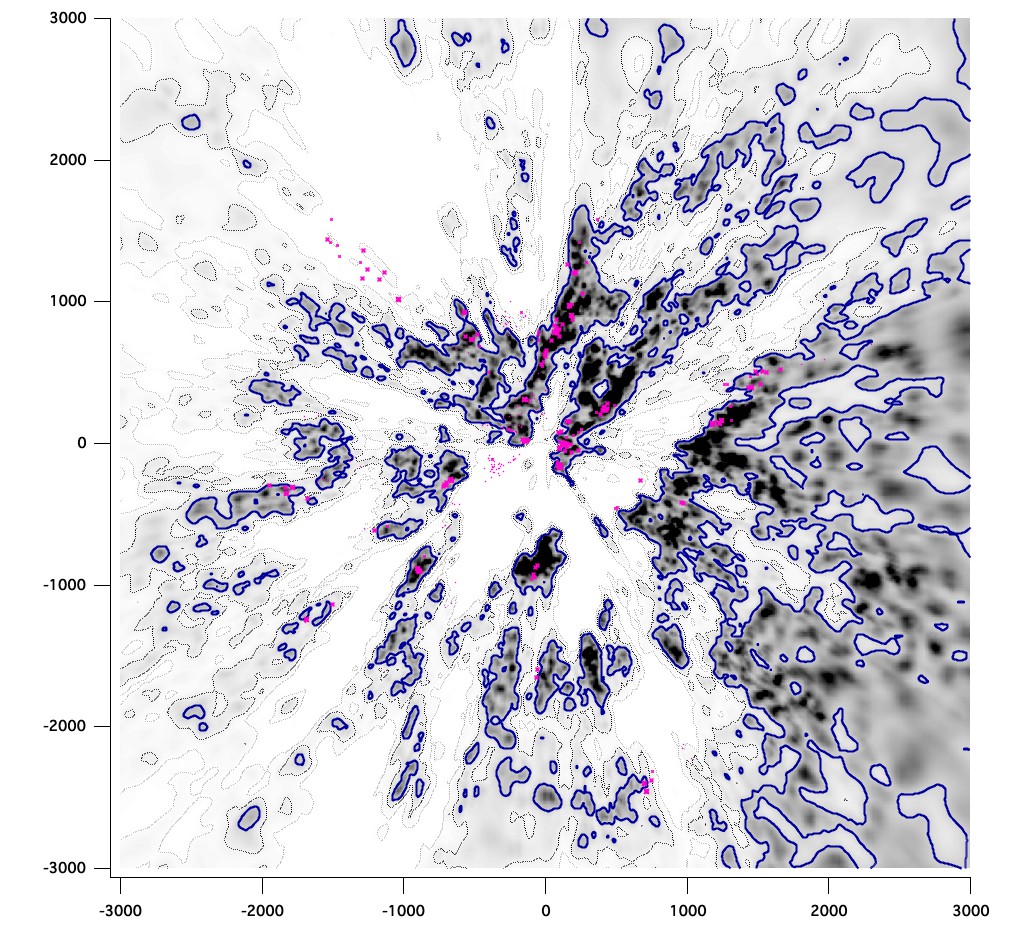}
 \caption{Dust in the mid-plane (black and white image) and projections onto the plane of molecular clouds from the \cite{Zucker20} catalogue (pink markers). The size of the marker decreases with the distance to the plane (maximum visible distance abs(Z)=100 pc.) Only a few objects from the catalogue fall outside dense dust structures: W3, W4, W5 (l=134 to 137 \fdeg. d=1-2 kpc), and IC1396 (l=99\fdeg, d=900 pc (but see text).}
  \label{mapandmcs}%
    \end{figure*}

\begin{figure*}
 \centering
  \includegraphics[width=0.88\hsize]{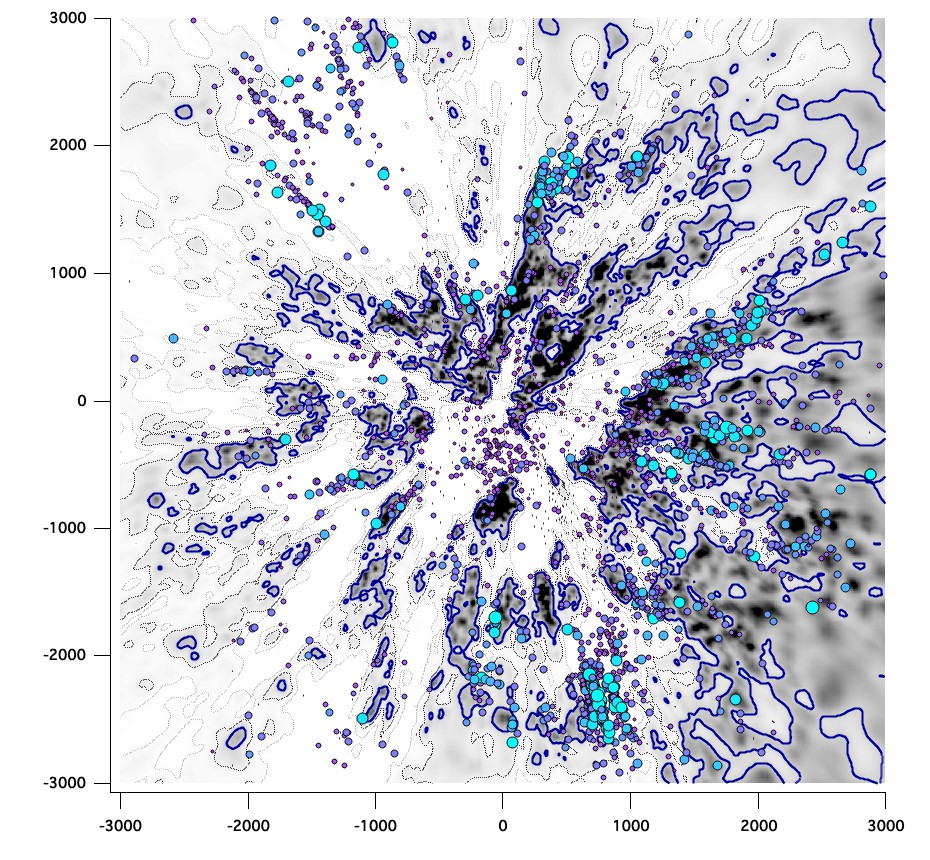}
 \caption{Massive O and B stars from the recent catalogue of \cite{Pantaleoni21}, superimposed on the dust extinction image in the horizontal plane containing the Sun. Only massive stars with identified stellar types, possessing EDR3 parallaxes, and located within 50 pc of the plane are represented. The colour scale and the size of the marker distinguish types from O3 (blue, large circles) to B9 (pink, small circles).}
  \label{Fig_OBstars}%
    \end{figure*}

It is important to compare dust cloud mapping based on distinct photometric data in order to detect potential differences. To do so, we have used the recent catalogue of molecular cloud (MC) distance assignments of \cite{Zucker20}. In this catalogue, molecular cloud locations are based on the PanSTARRS photometric data and {\it Gaia} DR2 parallaxes. For clouds within $\simeq$2 kpc, DR2 and EDR3 parallaxes slightly differ individually; however, the MC distance assignments are based on large numbers of targets, and errors are expected to be smoothed out in a large extent. Fig. \ref{mapandmcs} displays the projection onto the midplane of the $\geq$300 MCs from the catalogue, superimposed on the dust extinction in this plane. The figure shows a remarkable agreement, that is to say clouds are located within or are extremely close to the main dust concentrations deduced by inversion. There are only very few exceptions, namely W3 (IC 1795), W4(IC 805), W5 (IC 1848), and IC1396 (see  Fig. \ref{mapandmcs}). We note, however, that those particular clouds are much more diffuse than other objects of the catalogue, and that assigned locations of subsets of the first three are distributed over large distances. We also note some discrepancies between the assigned locations and distances of the dust {fronts} found using the \cite{Green19} online tool in the quoted directions. We attribute this variability to the faintness and large extent of these nebulae. Apart from these rare objects, the excellent agreement demonstrates that using different photometric bands does not introduce systematic trends in the distance assignments of the obscuring dust clouds. 

As said in the introduction, 3D maps are a general tool and comparisons with many types of observed quantities have a potential interest. We have selected one example, the comparison between the distribution of the clouds and cavities and the locations of massive early-type stars assembled in the recent  Alma catalogue of O and B stars from \cite{Pantaleoni21}. The catalogue is ideally suited for comparisons with our maps since the majority of the targets are located within 3 kpc. Before comparisons with the present dust distribution, and for consistency with the maps, we have searched for the {\it Gaia} EDR3 parallaxes of all objects from the catalogue and removed the very small number of objects without EDR3 parallaxes from the list. Fig. \ref{Fig_OBstars} shows, superimposed on the dust map in the mid-plane, the location of O, B stars identified as massive and located within 50 pc from the plane. A colour scale distinguishes the spectral types. It can be seen that, in general, O stars are located close to dense cloud complexes, which is in agreement with the classical scheme of star formation, according to which young massive star winds start to carve cavities in their parent cloud complexes immediately after their birth. There is, however, a difference in the second quadrant, characterised by a relatively large number of objects between 1 and 4 kpc, and the absence of dense dust structures. This behavior was already perceptible in our previous map; however, here the spatial coverage is improved and the mapping limitations can no longer be at the origin of this peculiarity. Such a difference deserves further studies, in particular it requires for potential effects of extinction law variability to be disentangled from actual different star formation episodes.

\begin{figure*}
 \centering
\includegraphics[width=0.68\hsize]{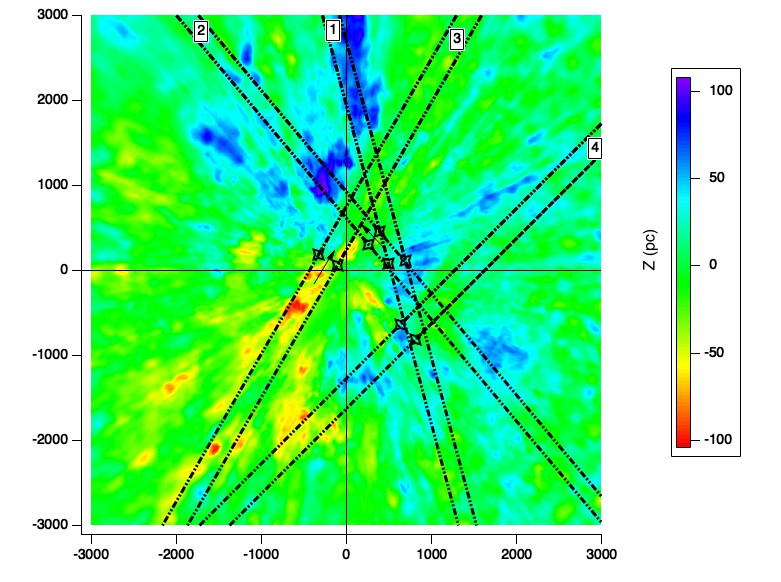}
\caption{Distance along the Z axis of the majority of the dust mass (see text for a definition). Distances are in parsecs and  positive towards the northern pole. In the third quadrant, most of the dust is below the mid-plane (but see text), while it is the opposite in the second and fourth quadrants. Also shown are the projections onto the mid-plane of the boundaries of the thick vertical slices from Fig. \ref{fig:ondul}, numbered from 1 to 4.}
  \label{fig:distfromplane}%
    \end{figure*}

\begin{figure*}
 \centering
 \includegraphics[width=0.68\hsize]{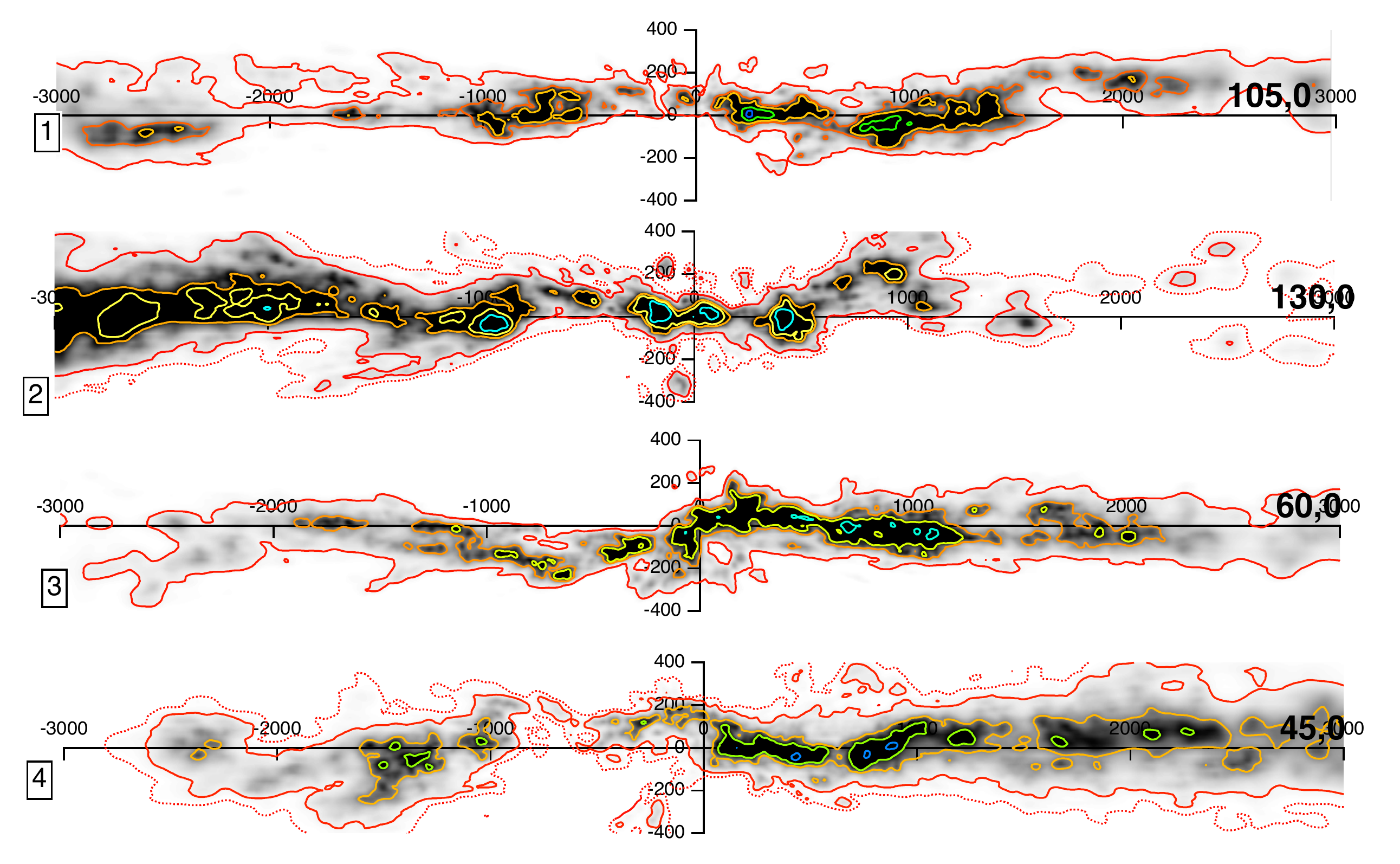}
 \caption{Examples of images of the dust in four vertical slabs. The orientations are indicated on the right, by means of the Galactic longitude of the corresponding extremity of the slab. The projections of the slabs onto the mid-plane are shown in Fig. \ref{fig:distfromplane}. The wavy pattern of the dust distribution with respect to the mid-plane is seen to be a general feature.}
  \label{fig:ondul}%
    \end{figure*}

Finally, a very peculiar and interesting aspect of this mapping is the vertical distribution of the dust. Fig. \ref{fig:distfromplane} displays, for each pair of X,Y coordinates, the quantity $Z_{mean}$ which represents the average Z coordinate of the dust (with the Z axis oriented from the southern to northern poles). More specifically, it was computed as the average value of the Z abscissa of each cell, weighted by the square of the extinction density in the cell,
\begin{equation}
 Z_{mean}= \frac {\int _{-400}^{+400} \rho(Z)^{2} Z dZ}{\int _{-400}^{+400} \rho(Z)^{2} dZ}
.\end{equation}
The figure shows a complex structure, dominated by a conspicuous negative trend in the third quadrant (dominant yellow and red colour in the figure) where most of the dust is 100-150 pc below the mid-plane. Conversely, a large fraction of positive values are found in the other quadrants, with most of the dust being 150-200 pc above the mid-plane in a number
of regions. In some specific areas, however, this average value has a limited significance due to the existence of two dust layers, such as in the anti-centre direction. In addition to this general trend, there are vertical wavy structures in many regions, with an amplitude in Z particularly large within 1 kpc from the Sun. This can be seen in Fig. \ref{fig:ondul} which shows examples of the dust density in four vertical slabs whose orientations, thicknesses, and intersections with the mid-plane are indicated in Fig. \ref{fig:distfromplane}. The quantity  represented in Fig. \ref{fig:ondul} is, for each Z value,  the integral of the extinction density along horizontal axes perpendicular to the slab orientation in the mid-plane. We have restricted the figure to four different slabs; however, the vertical oscillations of the dust layer appear to be a general feature. One of the vertical slabs (number three) contains the gas clouds forming the {Radcliffe wave} discussed by \cite{Alves20}, and it shows a remarkable sinusoidal pattern. It is beyond the scope of this work to discuss such structures in detail; however, we argue that they are likely connected with the {ripples} in the  3D velocity distribution of nearby stars recently discovered by \cite{Antoja18}. The spectacular spiral (or {snail-shaped}) pattern in the Z-V$_{Z}$ plane found by the authors is reproduced in state-of-the-art simulations of Milky-Way crossings by dwarf galaxies, but perturbations by the buckling bar and by arms also succeed in reproducing similar structures \citep[see, e.g. ][]{Hunt21,Khoperskov19}.  The period of the spiral along the Z direction can be derived from Fig. 1 of \cite{Antoja18} and is of the order of 300-350 pc. On the other hand, from Fig.\ 13, one can derive a mean vertical amplitude of the dust oscillations around the mid-plane of the same order visually, namely $\simeq$ 300-350 pc. We believe that this similarity is not accidental and that it is the signature of a common origin for both gas and stars perturbations. Future studies and, in particular, adjustments of self-consistent models for both stars and gas to stellar and interstellar matter distributions may help to infer the origin of both types of quasi-periodic structures. 
Fig. 2 from \cite{Antoja18} reveals another organised perturbation of the stellar velocities, namely ridges in the distribution of azimuthal velocities as a function of the Galactocentric radius.  Pursuing the search for correspondences between the dust distribution and the local stellar kinematics, we have used the figure to derive a mean spatial period of roughly 550 pc. However, we could not find clear signs of such a period along the X axis in the extinction density map. This is probably due to the complexity of the dust structures and especially the existence of two very different orientations of the main chains of dust clouds, a duality we already pointed out based on the DR2 maps: a l$\simeq$80\fdeg direction similar to the spiral arm orientation and, second, a l$\simeq$45\fdeg orientation illustrated by the 'split' and a large fraction of Sagittarius (see Fig. 10 and 11 from \cite{Lallement19}). Such a complex pattern suggests the existence of past perturbations, other than the one having generated the {snail-shaped} pattern. 

\section{Conclusions and perspectives}

Thanks to the new version of the {\it Gaia} data (EDR3), and in particular to the more accurate parallaxes and more homogeneous photometry, we have updated our calculations of the 3D extinction density distribution in the local interstellar medium at $\lambda$=5500\AA.  The computational volume is a 6 kpc x 6 kpc x 800 pc slab with the Sun at the centre. We have also provided error estimates on both the extinction densities (or differential extinction, in $mag\cdot pc^{-1}$) and on the integrated extinctions on the lines of sight (from the Sun to any location within the volume). The maps are downloadable in their entirety (3D distribution of 5pcx5pcx5pc voxels) and/or available for inspection with several tools on the EXPLORE website\footnote{https://explore-platform.eu}. 
In particular, it is possible to download integrated extinction profiles for series of directions captured in a single file, as well as dust distribution images in selected planes. 

The main improvements to the maps are improved cloud reconstruction quality, increased distances at which clouds are reconstructed, and significantly improved local cavity reconstruction. 
The new map confirms the highly complex 3D distribution of dust at 3-4 kpc, an unexpected structure before {\it Gaia}. The 45\fdeg \ pitch angle cloud chain nicknamed 'split' and the huge dust-free 'avenue' between 'split' and Sagittarius are clearly visible. The undulations of the dust sheet around the plane are extensive.  The average vertical amplitude of these oscillations around the mid-plane is of the order of $\Delta$Z=300 pc, which is similar to the vertical period found in the 3D velocity distribution of nearby stars by \cite{Antoja18}, namely the spectacular spiral (or {>snail}) pattern of azimuthal (tangential) and radial velocities in the Z-V$_{Z}$ plane. We suggest a common origin for these two phenomena. If this is the case, the grain and gas distributions could be associated with the kinetic 3D tomography of the stars in order to help clarify their origin. 

 With the low-resolution blue and red spectrographs on board the satellite, future {\it Gaia} data releases will provide a much larger number of reliable extinction measurements. The dual hierarchical method has proven to be well suited to the inversion of large input data sets due to its low computational time and will be well adapted to these data. In the future, one may expect much improved dust distributions, ideally accompanied by co-spatial gas kinematics. Such results may enter self-consistent models of stars and gas to help decipher the origin of the present-day Galactic structure and the epochs and impacts of disk traversals and bar and arm disruptions. In this regard, the construction of fully self-consistent distributions of extinction and stars \citep{Babusiaux20, Hottier20} will be a crucial improvement.


%

\begin{acknowledgements}

We thank our anonymous referee for useful comments.\\
J.L.V. and N.L.J.C. acknowledge support from the EXPLORE project. EXPLORE has received funding from the European Union’s Horizon 2020 research and innovation programme under grant agreement No 101004214.
J.L.V. acknowledges support from the THETA Observatoire des Sciences de l'Univers in Besan\c{c}on.\\
This work has made use of data from the European Space Agency (ESA) mission Gaia (https://www.cosmos.esa.int/gaia), processed by the Gaia Data Processing and Analysis Consortium (DPAC, https://www.cosmos.esa.int/web/gaia/dpac/consortium). Funding for the DPAC has been provided by national institutions, in particular the institutions participating in the Gaia Multilateral
Agreement. This work also makes use of
data products from the 2MASS, which is a joint project of
the University of Massachusetts and the Infrared Processing
and Analysis Center/California Institute of Technology,
funded by the National Aeronautics and Space Administration
and the National Science Foundation.\\
This research has made use of the SIMBAD database, operated at CDS, Strasbourg, France.
\end{acknowledgements}

%
\bibliographystyle{aa} 
\bibliography{mybib} 
%
\end{document}